\documentclass[nopubldata]{lpr2}
\usepackage{hyperref}
\usepackage{epstopdf}
\hypersetup{%
  colorlinks,
  plainpages=false,
  pdfpagelabels,
  breaklinks,
  pdfview=Fit,
  bookmarksopen,
  bookmarksnumbered,
  linkcolor=black,
  anchorcolor=black,
  citecolor=black,
  filecolor=black,
  urlcolor=LPRred
  }%
\graphicspath{{figs/}}
\begin{document}
%
\titlefigure{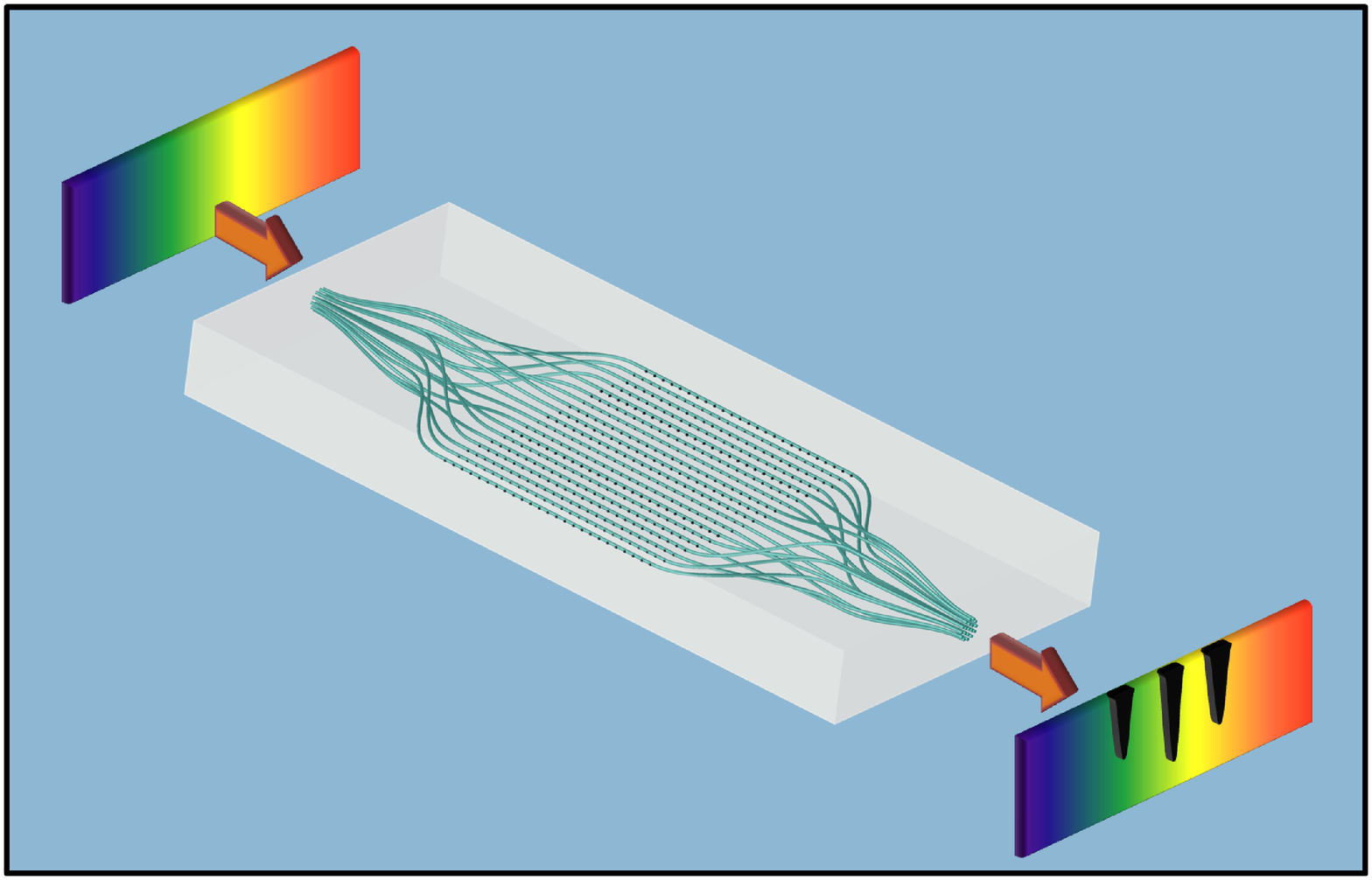}

\abstract{The first demonstration of narrowband spectral filtering of multimode light on a 3D integrated photonic chip using photonic lanterns and waveguide Bragg gratings is reported. The photonic lanterns with multi-notch waveguide Bragg gratings were fabricated using the femtosecond direct-write technique in boro-aluminosilicate glass (Corning, Eagle 2000). Transmission dips of up to 5 dB were measured in both photonic lanterns and reference single-mode waveguides with 10.4-mm-long gratings. The result demonstrates efficient and symmetrical performance of each of the gratings in the photonic lantern. Such devices will be beneficial to space-division multiplexed communication systems as well as for units for astronomical instrumentation for suppression of the atmospheric telluric emission from OH lines.}

\title{Multiband processing of multimode light: combining \\3D photonic lanterns with waveguide Bragg gratings}
%
\titlerunning{Multiband processing of multimode light}
\author{Izabela Spaleniak,\inst{1,2,*} Simon Gross,\inst{1,3} Nemanja Jovanovic,\inst{4}  Robert J. Williams, \inst{1}\\ Jon S. Lawrence, \inst{1,2,5} Michael J. Ireland, \inst{1,2,5} and Michael J. Withford\inst{1,2,3}}
%
\authorrunning{I. Spaleniak et al.}
%
\institute{%
   MQ Photonics Research Centre, Dept. of Physics and Astronomy, Macquarie University, NSW 2109, Australia\\
         \and
           Macquarie University Research Centre in Astronomy, Astrophysics \& Astrophotonics, Dept. of Physics and Astronomy, Macquarie University, NSW 2109, Australia \\
	\and
	Centre for Ultrahigh Bandwidth Devices for Optical Systems (CUDOS)\\
	\and
	National Astronomical Observatory of Japan, Subaru Telescope, 650 N. A'Ohoku Place, Hilo, Hawaii, 96720, U.S.A.\\
         \and
	Australian Astronomical Observatory (AAO), 105 Delhi Rd, North Ryde NSW 2113, Australia\\
	 }
%
\mail{\textsuperscript{*}\,Corresponding author: e-mail: izabela.spaleniak@mq.edu.au}
%
%
\maketitle

\noindent The application of photonic technologies to astronomical instrumentation is a burgeoning new field which offers new dimensions in design flexibility, as well as dramatic size and cost reductions for increasingly large telescope projects. One critical development is the photonic lantern, which provides an efficient interface between multimode (MM) ''light-bucket'' fibres, and high fidelity, single-mode (SM) photonic devices. A photonic lantern is a tapered transition that converts light from a MM input into a number of SM outputs \cite{Saval}. The initial motivation for these devices came from the field of astrophotonics \cite{Joss}, but more recently photonic lanterns were also proposed for space-division multiplexed optical communication across MM fibers \cite{Fontaine}.

Since light from a MM fiber is converted into many SM fibers, photonic lanterns enable the use of SM photonic components, such as fiber Bragg gratings for narrow-band filtering of MM light. This idea was realized in an astronomical instrument called GNOSIS \cite{Trinh} to suppress the earth's atmospheric OH emission lines coming from the de-excitation of OH molecules from the starlight signal before it was injected into a spectrograph. In that case the instrument is composed of seven 19-channel fiber back-to-back (MM-SM-MM) photonic lanterns, where each of the 19 SM channels is equipped with identical multi-notch fiber Bragg gratings that efficiently suppress the 103 brightest OH lines in the near-infrared part of the spectrum. This requires that all fiber Bragg gratings are exactly identical \cite{Saval,Trinh}, so that the same part of the spectrum is always reflected, independent on how the light is spread from the MM input across the SM channels.

As a result of being fully made out of fibers, the instrument is large and fragile, and thus delicate to handle. Therefore it requires careful packaging in order to avoid any stress or temperature variations across the fiber Bragg gratings, which would deteriorate the instrument's performance. Furthermore, realizing fiber based lanterns becomes increasingly more difficult with increasing number of SM outputs. In order to improve the robustness of the instrument, efforts have been made towards using multicore optical fibers \cite{Birks}. However, the inscription of Bragg gratings uniformly across all cores for a multicore fiber has proven to be challenging \cite{Birks}. Alternatively, the integration of a lantern onto a photonic chip was proposed \cite{Thomson2009} and then demonstrated \cite{Thomson2011} by using ultrafast laser inscription, albeit without spectral filtering. 

In our previous work we used a femtosecond laser to inscribe a series of devices in order to optimize the MM waveguide input \cite{Jovanovic} and the photonic lantern \cite{Spaleniak} design itself. In this Letter we demonstrate fully integrated photonic lanterns with incorporated waveguide Bragg gratings (WBGs) to provide spectral filtering and discuss potential astronomical applications.
\begin{figure}
\centerline{\includegraphics[width=7.5cm]{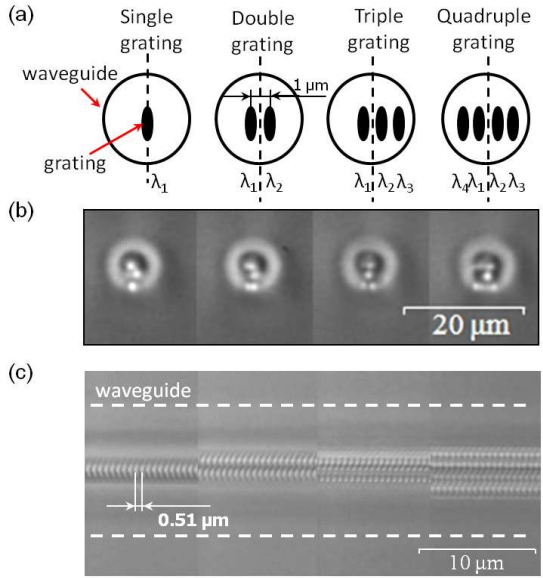}}
\caption{(a) Sketch of the cross-section  of the waveguides (circles) with the gratings (elipses) with indicated position of the center of the waveguide (dotted line); $\lambda_{1}$~=~1545~nm, $\lambda_{2}$~=~1552 nm, $\lambda_{3}$~=~1559 nm,  $\lambda_{4}$~=~1563 nm; microscope images of the SM waveguides seen from the front (b) and top (c) with one, two, three and four gratings.} 
\label{schematics}
\end{figure}

Femtosecond laser written WBGs have been demonstrated using various techniques, such as single axial point-by-point \cite{Marshall,Gross} or a square-wave modulated pulse train \cite{Zhang}. The strength of the Bragg resonance depends on the length of the grating and the degree of mode overlap between the waveguide and the grating modifications. While the single-step fabrication using a square-wave modulated pulse train results in better mode overlap, thus stronger gratings, the sequential single axial point-by-point approach allows for placing multiple gratings in a single waveguide \cite{Koutsides,Thomas2011}, because the waveguide is written first and followed by the insription of the periodic grating modifications in a second step. The strength of the grating can then be readily enhanced by increasing the length of the WBG.

The fabricated chip contains multiple devices, each consisting of back-to-back lanterns with a horizontal array of straight waveguides in the middle as illustrated in the abstract figure. Each photonic lantern is composed of 19 individual SM waveguides which are brought close to each other to form the MM waveguide input \cite{Jovanovic}. These 19 waveguides then fan out into a horizontal array of straight waveguides. Each SM waveguide either contains one, two, three or four gratings with different Bragg resonances placed next to each other in the SM waveguide core (Fig.~\ref{schematics}(a)). The total chip length of 39~mm includes: 2-mm-long MM waveguides at both ends, 12-mm-long transitions from MM waveguides to the horizontal array of straight SM waveguides and the 10.4-mm-long straight WBG section in the middle.

The devices were inscribed into a boro-aluminosilicate glass (Corning Eagle2000) using an ultrafast Ti:sapphire oscillator (Femtolasers FEMTOSOURCE XL 500, 800~nm, 5.1~MHz, 550~nJ, $<$~50 fs). A 100$\times$~1.25 numerical aperture (NA) oil immersion objective (Zeiss N-Achroplan) was used to focus the circular polarized laser beam 300~$\mu$m below the sample's top surface. The photonic lantern structures were written in the cumulative-heating regime with the laser's repetition rate of 5.1~MHz at a pulse energy of 35~nJ and a translation speed of 750~mm/min. The strong heat diffusion in the cumulative heating regime inhibits the fabrication of micrometer- or sub-micrometer sized structures, which are required for low-order Bragg gratings. In a second step after the waveguide inscription the laser repetition rate was reduced to 52.022~kHz with an external electro-optic pulse picker to place point-by-point modifications for the gratings into each of the SM waveguides. Due to the high focusing NA, the point-by-point features are only $\sim$1$\times$4~$\mu$m in cross section, multiple gratings can then be placed side-by-side spaced by 1~$\mu$m within the $\sim$10~$\mu$m diameter SM waveguide, as shown in Fig.~\ref{schematics} (a, b, c). The grating inscription pulse energy was optimized for maximum coupling strength: using 125~nJ pulses we achieved coupling strength $\kappa$~=~$\sim$1.2~cm$^{-1}$. The period and therefore the Bragg wavelength ($\lambda_{B}$) of the gratings was tuned by changing the translation speed of the stages according to $v_{trans}=\lambda_{B}\cdot f_{rep}/(2n_{eff})$, where $v_{trans}$ is the stages translation velocity, $f_{rep}$ repetition rate and $n_{eff}$ is the effective refractive index. In this fashion four photonic lantern structures were written with single ($\lambda_{B}$~=~1545~nm), double ($\lambda_{B}$~=~1545~nm \& 1552~nm), triple ($\lambda_{B}$~=~1545~nm, 1552~nm \& 1559~nm) and quadruple ($\lambda_{B}$~=~1545~nm, 1552~nm, 1559~nm \& 1563~nm) WBGs.  The typical translation speed was 1620~mm/min (27~mm/s), hence the fabrication process of the triple grating photonic lantern took $\sim$~6 minutes. 

In order to quantify the spectral performance of the photonic lanterns, identical WBGs were inscribed into isolated SM waveguides for reference purposes. The devices were spectrally characterized using a swept wavelength system (SWS15100, JDS Uniphase) with 3~pm resolution. To ensure even illumination of the MM input ($\sim$~50~$\mu$m) the light was free-space coupled into the chip with a focal spot size in the order of the input MM waveguide diameter. The transmitted light was collected with a MM fiber and fed into the detector of the swept wavelength system.

\begin{figure}
\centerline{\includegraphics[width=7.8cm]{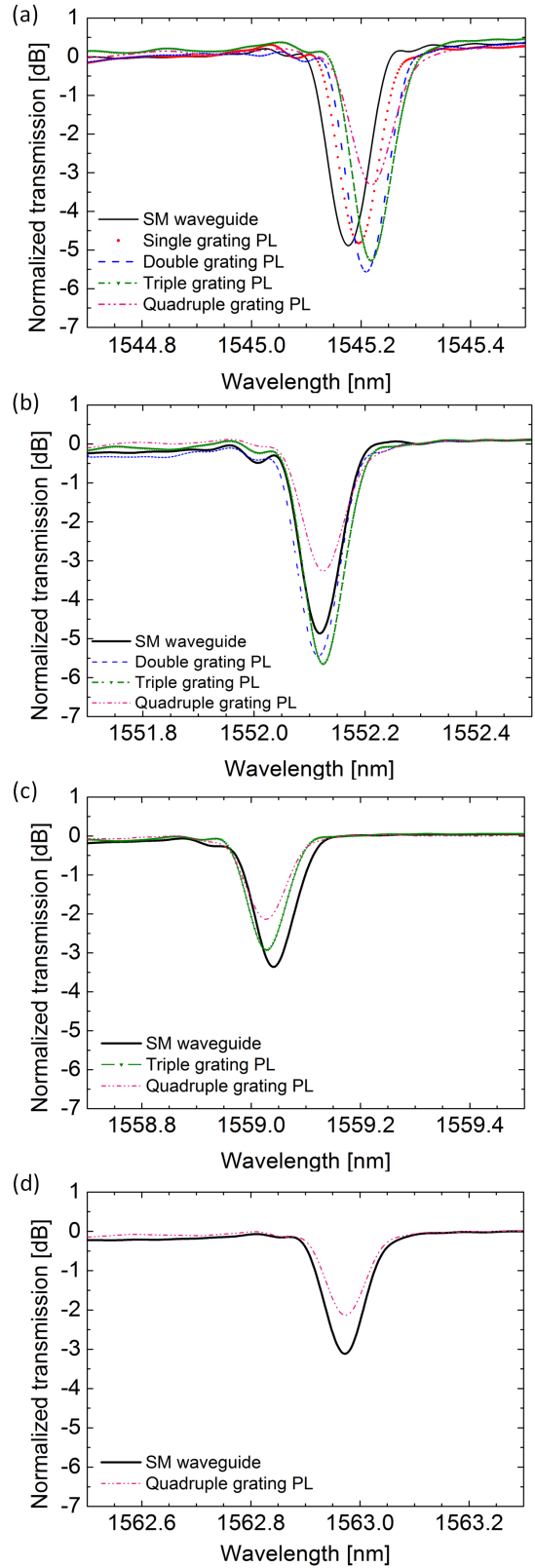}}
\caption{Normalized transmission spectra of the photonic lanterns (PL) and SM waveguides at (a) 1545~nm, (b) 1552~nm, (c) 1559~nm, and (d) 1563~nm.}
\label{graphs}
\end{figure}

Figure~\ref{graphs} shows the normalized transmission spectra of the 19-channel photonic lanterns and the reference SM waveguide with 10.4-mm-long first order gratings at 1545 nm (Fig.~\ref{graphs}(a)), 1552 nm (Fig.~\ref{graphs}(b)), 1559 nm (Fig.~\ref{graphs}(c)) and 1563 nm (Fig.~\ref{graphs}(d)). Table \ref{table1} summarizes the transmission dip depths of the spectral lines. As seen from the figures and the table, the gratings in the photonic lanterns exhibit a very similar performance to the gratings in the reference SM waveguide. This result proves an efficient and symmetrical performance of each of the gratings in the photonic lantern, which demonstrates that the femtosecond direct-write process provides excellent reproducibility. In contrast, the fiber Bragg gratings inscribed into multicore fibers \cite{Birks} suffer from non-uniform strength and resonance wavelengths across the different cores. This was attributed to problems with achieving a uniform illumination of all the cores through the phase mask during the fabrication process.

The strongest resonances for our gratings are obtained from the modifications located closest to the center of the core because of the strongest overlap with the guided mode (i.e. 1545~nm and 1552~nm as seen in Fig.~\ref{schematics}(a)). However an optimized arrangement of the grating modifications could be used to obtain more even resonances. For the layout of modifications in the waveguides used in this work, the triple grating photonic lantern offered the best performance as it exhibited strong transmission dips simultaneously at multiple wavelengths. Slight Bragg wavelength shifts are present (Fig.~\ref{graphs}(a)). This is consistent with the expected change in the guided mode's effective index  $n_{eff}$ as more gratings are placed within the waveguide cores. The FWHM of the grating resonances are in the order of 0.1~nm, and as such sufficiently narrow to be used for OH emission filtering \cite{Trinh}. The number of filter lines could be increased by inscribing complex aperiodic \cite{Joss2008}, amplitude- or phase-sampled WBGs \cite{Marshall2010}.
\begin{table}
  \centering
  \caption{Summary of the depths of the gratings' transmission dips in the photonic lanterns (PL) and SM waveguide (SM WG).}\begin{tabular}{cccccc} \\ \hline
 \label{table1}
   &  \multicolumn{4}{c} {\textbf{Dip depth [dB]}} \\   [-0.5ex] \cline{2-5} 
     \raisebox{1.5ex}{\textbf{Structure}} &  \multicolumn{4}{c} {\textbf{Wavelength [nm]}} \\  [-1ex]
     \raisebox{1.5ex}{\textbf{type}} & 1545 & 1552 & 1559 & 1563 \\  \hline 
    Single grating (PL) & 4.89 &  &  &  \\
    Double grating (PL)  & 5.49 & 5.38 &  & \\
    Triple grating (PL)  & 5.12 & 5.60 & 2.87 & \\
    Quadruple grating (PL)  & 3.25 & 3.26 & 2.03 & 2.14 \\
    SM WG & 4.86 & 4.42 & 3.30 & 3.12 \\ \hline
  \end{tabular}
\end{table}

\begin{figure}[htb]
\centerline{\includegraphics[width=8.5cm]{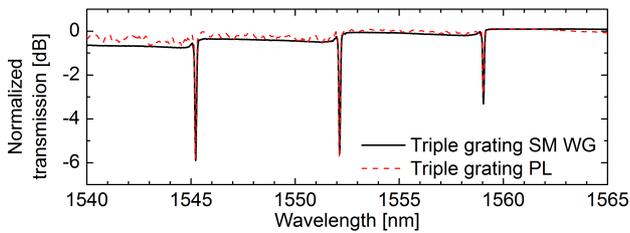}}
\caption{Normalized transmission spectra of the triple grating SM waveguide and triple grating photonic lantern. The transmission losses on the short wavelength side are caused by coupling into radiation modes.}
\label{triple}
\end{figure}

Figure~\ref{triple} presents normalized transmission spectra of multiple notches for the triple grating photonic lantern and reference triple grating SM waveguide. Transmission losses of $\approx$~0.3$\--$0.4~dB/grating/cm are apparent on the short wavelength side of each resonance due to coupling into the continuum of radiation modes. These losses can be reduced by better matching of the grating and waveguide spatial profiles or using the femtosecond phase mask technique \cite{Thomas}. The total normalized throughputs of $>$ 95\% have been reported for femtosecond laser written photonics lanterns \cite{Spaleniak}.

The calculated coupling strength coefficients for the triple photonic lantern are $\kappa_{1545nm}$~=~1.15~cm$^{-1}$, $\kappa_{1552nm}$~= 1.21~cm$^{-1}$ and $\kappa_{1559nm}$~=~0.83~cm$^{-1}$.  Therefore the grating of 20~mm would already give 15~dB strong notches at 1545~nm and 1552~nm and 9~dB at 1559~nm. The required strength for OH suppression is of the order of 20-30~dB, which can be obtained with a grating of about 35~mm. This can be easily achieved with longer glass samples, as with our current fabrication system we can fabricate photonic structures as long as 100 mm.
The astronomical application of fiber Bragg gratings has so far involved filtering atmospheric OH lines before injection into a spectrograph \cite{Trinh}. The frequently used alternative to this is post-dispersion suppression, which can be as simple as masking out pixels in a medium-resolution (R~$>$~3000) spectrograph where bright OH lines appear. Post-dispersion suppression has the disadvantage that it cannot suppress the energy from the OH lines scattered by imperfections in the spectrograph's grating, leaving a residual background that is higher than scattered zodiacal light. However, the merit of OH suppression is not simply measured by the signal-to-noise of a single background-limited observation. Many modern instruments (e.g. MOSFIRE \cite{McLean}) have a figure of merit that involves optimizing low background, high efficiency and highly multiplexed platform within a fixed cost envelope. Furthermore, as instrumentation moves systematically from seeing-limited instrumentation to diffraction-limited instrumentation (e.g. GMTIFS \cite{McGregor}), the back ground is lower and there is a role for OH suppression at only the 20~dB level. 

One obvious vision for the structures presented here is future development of long, complex gratings used pre-dispersion in an astronomical spectrograph. An alternative vision is to use the spectral filters for only a few OH lines post-dispersion in a low spectral resolution multi-object or integral field unit spectrograph. The high speed of direct-write (more than 20 mm/s in this paper) means that it is possible that after a spectrum is dispersed, a very large number of structures such as that in the abstract figure could be place on top of each other (or rotated 90 degrees about the injection axis and placed side by side). This would essentially mean that a separate photonic lantern device is tailored for every spectral resolution element of each object observed. For this application, complex gratings  \emph{are not required}, with only a few strong lines in the $\sim$1\,nm spectral bandwidth represented by each pixel. This also has a great advantage over direct geometric OH suppression (e.g. an image-plane mask) because the spectrograph itself is only low-resolution -- the high dispersion part of the instrument would be fully contained in the integrated optics fiber Bragg gratings.

We have demonstrated an integrated photonic lantern structure with integrated spectral filters. The spectral characteristics of the WBGs match those of existing astrohotonics devices but with the benefit that ultrafast laser inscription creates integrated devices which are inherently robust and offer flexibility in regards of the number of the waveguides and their arrangement. This technology enables the miniaturization of the existing astrophotonic components by a few orders of magnitude as well as realizing integrated add/drop filters in multiple-input multiple-output systems \cite{Fontaine}.


\begin{acknowledgement}
This research was supported by the Australian Research Council Centre of Excellence for Ultrahighbandwidth Devices for Optical Systems (project no. CE110001018) and the OptoFab node of the Australian National Fabrication Facility. I. Spaleniak acknowledges the support of the iMQRES scholarship and AAO top-up scholarship.
\end{acknowledgement}

\begin{biographies}
\end{biographies}


\end{document}